\documentclass[12pt]{article}
\usepackage{graphics}
\newcommand{\myand}{}
\begin{document}
\begin{center}
{\bf \LARGE Central Pb+Pb Collisions at 158 A GeV/c Studied by $\pi^-\pi^-$ Interferometry}
\end{center}
\vspace*{0.7cm}
{\small \noindent
M.M.~Aggarwal,$^{1}$ \myand
A.~Agnihotri,$^{2}$ \myand
Z.~Ahammed,$^{3}$ \myand
A.L.S.~Angelis,$^{4}$ \myand
V.~Antonenko,$^{5}$ \myand
V.~Arefiev,$^{6}$ \myand
V.~Astakhov,$^{6}$ \myand
V.~Avdeitchikov,$^{6}$ \myand
T.C.~Awes,$^{7}$ \myand
P.V.K.S.~Baba,$^{8}$
% ...
}
\\ \vspace*{-0.1cm}
\begin{center}
{\normalsize (WA98 Collaboration)}
\end{center}
\vspace*{0.2cm}
{\small
{$^{1}$~University of Panjab, Chandigarh 160014, India} \\
{$^{2}$~University of Rajasthan, Jaipur 302004, Rajasthan, India} \\
{$^{3}$~Variable Energy Cyclotron Centre,  Calcutta 700 064, India} \\
{$^{4}$~University of Geneva, CH-1211 Geneva 4, Switzerland} \\
{$^{5}$~RRC Kurchatov Institute, RU-123182 Moscow, Russia} \\
{$^{6}$~Joint Institute for Nuclear Research, RU-141980 Dubna, Russia} \\
{$^{7}$~Oak Ridge National Laboratory, Oak Ridge, Tennessee 37831-6372, USA} \\
{$^{8}$~University of Jammu, Jammu 180001, India} \\
{$^{9}$~University of M{\"u}nster, D-48149 M{\"u}nster, Germany} \\
{$^{10}$~SUBATECH, Ecole des Mines, Nantes, France} \\
{$^{11}$~Gesellschaft f{\"u}r Schwerionenforschung (GSI), D-64220 Darmstadt, Germany} \\
{$^{12}$~Universiteit Utrecht/NIKHEF, NL-3508 TA Utrecht, The Netherlands} \\
{$^{13}$~Lund University, SE-221 00 Lund, Sweden} \\
{$^{14}$~University of Tsukuba, Ibaraki 305, Japan} \\
{$^{15}$~Nuclear Physics Institute, CZ-250 68 Rez, Czech Rep.} \\
{$^{16}$~KVI, University of Groningen, NL-9747 AA Groningen, The Netherlands} \\
{$^{17}$~Institute for Nuclear Studies, 00-681 Warsaw, Poland} \\
{$^{18}$~MIT, Cambridge, MA 02139, USA} \\
{$^{19}$~Institute of Physics, 751-005  Bhubaneswar, India} \\
{$^{20}$~University of Tennessee, Knoxville, Tennessee 37966, USA} \\
}
%
\normalsize
\abstract{
Two-particle correlations have been measured for identified
$\pi^-$ from central 158 AGeV Pb+Pb
collisions and fitted radii of about 7 fm in all dimensions have been obtained.
A multi-dimensional study of the radii as a function of $k_T$
is presented, including a full correction for the
resolution effects of the apparatus.
The cross term $R^2_{out-long}$ of the standard fit in the Longitudinally
CoMoving System (LCMS) and the $v_L$
parameter of the generalised Yano-Koonin fit are compatible with
0, suggesting that the source undergoes a
boost invariant expansion.
\\
The shapes of the correlation functions in $Q_{inv}$ and
$Q_{space}=\sqrt{Q^2_x+Q^2_y+Q^2_z}$ have been analyzed in detail.
They are not Gaussian but better represented by exponentials.
As a consequence, fitting Gaussians to these correlation 
functions may produce different radii depending on the 
acceptance of the experimental setup used for the measurement.
} 
%
%

\section{Introduction}

The study of Bose-Einstein correlations between pairs of 
identical hadrons is an essential tool to obtain information
on the space-time evolution of the extended hadron sources created in
heavy ion collisions \cite{BE}. In particular, a strong correlation
between the momenta and the space-time production points of the
particles suggests expanding sources as predicted by hydrodynamic
models \cite{Csorg}. The dynamical evolution of such systems can then be
studied with interferometry via selection on the transverse momenta and
rapidity of the correlated particle pairs.

In this paper, we present the analysis of two-particle 
correlations of identified $\pi^-$ measured in the WA98 experiment
for central 158 AGeV Pb+Pb collisions at the CERN SPS.

\section{Experimental setup and data processing}

\begin{figure*}
\rotatebox{270}{
 \resizebox{0.75\textwidth}{!}{%
  \includegraphics{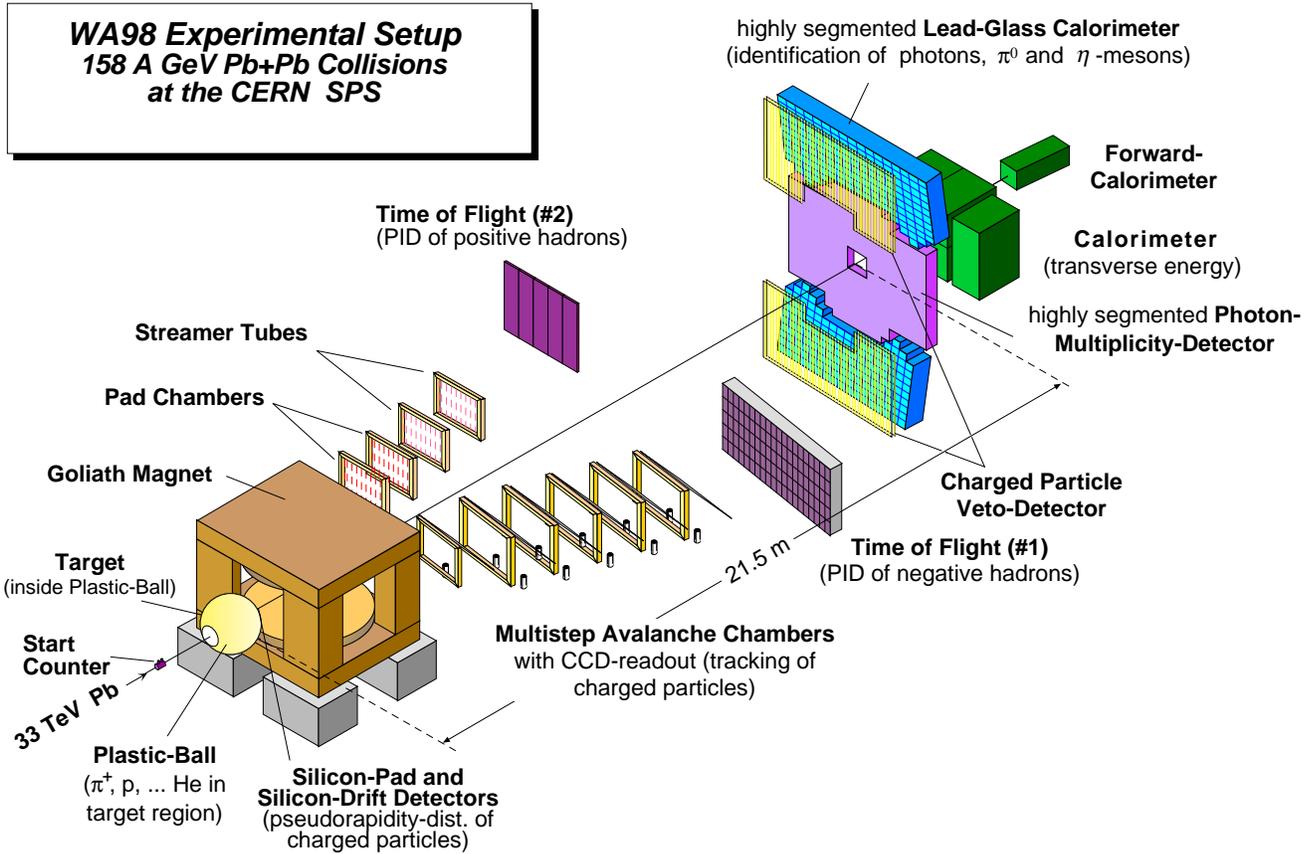}
}
}
\vspace*{0cm}       
\caption{Experimental setup.}
\label{fig:figure1}       
\end{figure*}

The WA98 experiment shown in Fig.~\ref{fig:figure1} combined large
acceptance photon
detectors with a two arm charged particle tracking spectrometer.
The incident 158 AGeV Pb beam
interacted with a Pb target near the entrance of a large
dipole magnet.
Non-interacting beam nuclei, or beam fragments were detected in a forward
calorimeter located at zero degree.
A mid-rapidity calorimeter measured the total transverse energy
in the rapidity region 3.2 $\leq$ $\eta$ $\leq$ 5.4,
which was also used in the trigger for online centrality selection.
The Plastic-Ball calorimeter measured the fragmentation of the
target, and silicon detectors were used to measure the charged
particle multiplicity.
The photon detectors consisted of a large area photon multiplicity
detector and a high granularity lead-glass calorimeter
for single photon, $\pi^0$, and $\eta$ physics \cite{Peitz}.
 
The charged particle spectrometer made use of a 1.6 Tm
dipole magnet with a 2.4$\times$1.6 m$^2$ air gap for
magnetic deflection of the charged particles in the horizontal
plane. The results presented in this paper are taken
from the 1995 WA98 data set obtained with the negative particle
tracking arm of the charged particle spectrometer. The second tracking
arm was added to the spectrometer in 1996 to measure positive
particles \cite{LUND}.
The first tracking arm consisted of six multistep avalanche
chambers with optical readout \cite{MSAC} located downstream of the magnet.
The active area of the first chamber was 1.2$\times$0.8 m$^2$,
while that of the other five was 1.6$\times$1.2 m$^2$.
The chambers contained a photoemissive vapour (TEA) which produced
UV photons along the path of traversal of the charged particles.
These were converted
into visible light via wavelength shifter plates. On
exit the light was reflected by mirrors at 45$^{\circ}$ to
CCD cameras equipped with two image intensifiers. Each pixel
of a CCD viewed a 3.1$\times$3.1 mm$^2$ area of a chamber.
In addition, a 4$\times$1.9 m$^2$ Time of Flight wall positioned
behind the chambers at a distance of 16.5 m from the target allowed for
particle identification with a time resolution better than 120 ps.

Fig.~\ref{fig:figure2} shows the Monte Carlo generated $p_T$-rapidity
acceptance for $\pi^-$.
The acceptance ranges from $y$=2.1 to 3.1 with an average at 2.70.
The momentum resolution of the spectrometer was $\Delta p/p$=0.005
at $p$=1.5 GeV/c, resulting in an average
accuracy better than or equal to 10 MeV/c
for all the Q variables used in the interferometry analysis and
defined in section 5: ${\sigma}(Q_{inv})$=7 MeV/c,
${\sigma}(Q_{TO})$=10 MeV/c, ${\sigma}(Q_{TS})$=5 MeV/c,
${\sigma}(Q_L)$=3 MeV/c, ${\sigma}(Q_{T})$=8 MeV/c, ${\sigma}(Q_{0})$=5 MeV/c.
 
The analysis of the complete 1995 data set is presented here. These data
have been taken with the most central triggers corresponding to about 10\%
of the minimum bias cross section of 6190 mb.
Severe track quality cuts were applied
at the expense of statistics
resulting in final samples of 4.2$\times10^6$ $\pi^-$ for the
correlation analysis and 4.6$\times10^5$ $\pi^-$ for the single particle
spectrum.

\section{Single particle spectra}

\begin{figure}[!h]
\hspace*{4.0cm}
\resizebox{0.5\textwidth}{!}{%
  \includegraphics{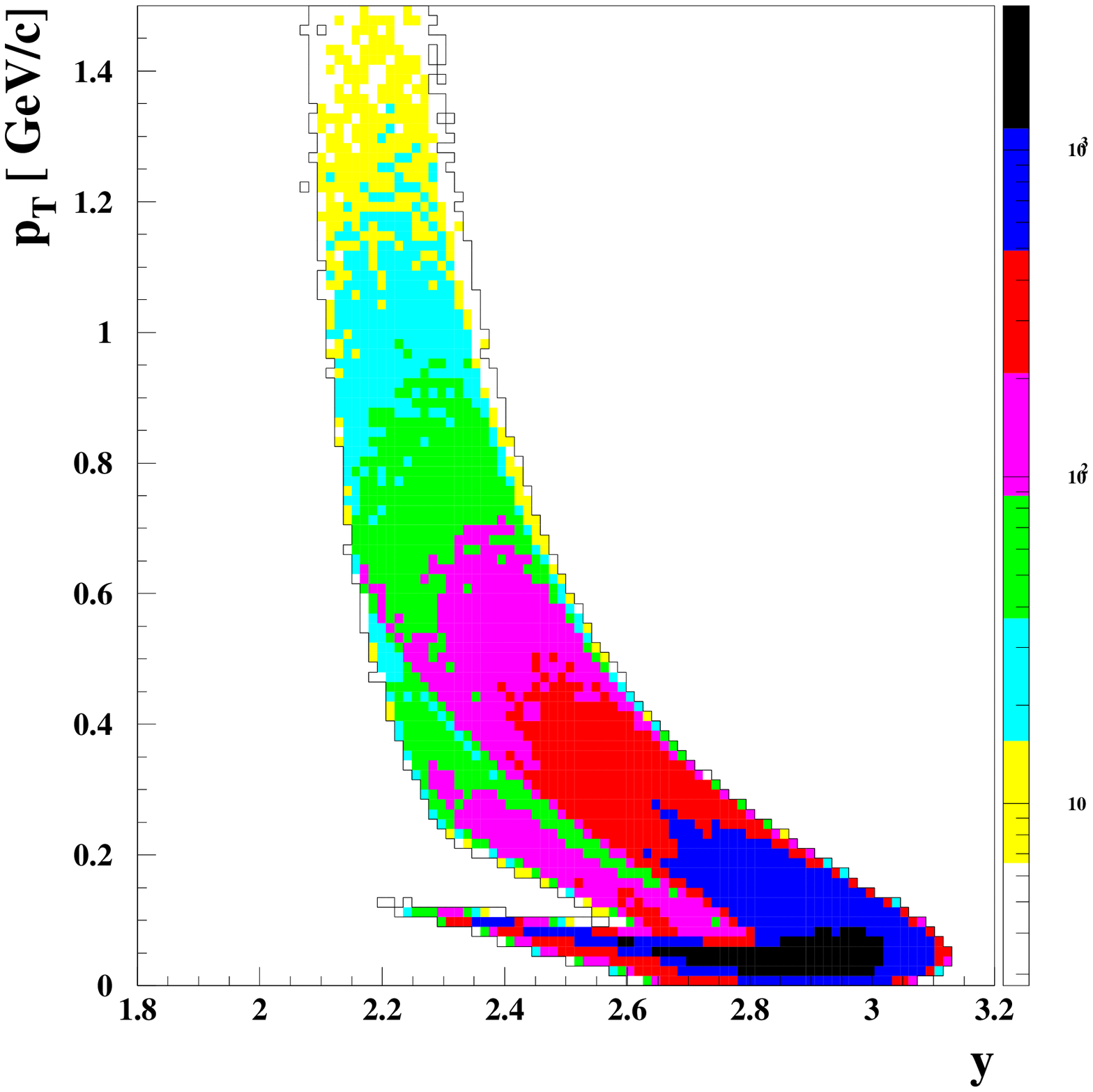}
}
\caption{$p_T$-rapidity acceptance.}
\label{fig:figure2} 
\hspace*{4.3cm}
\resizebox{0.47\textwidth}{!}{%
  \includegraphics{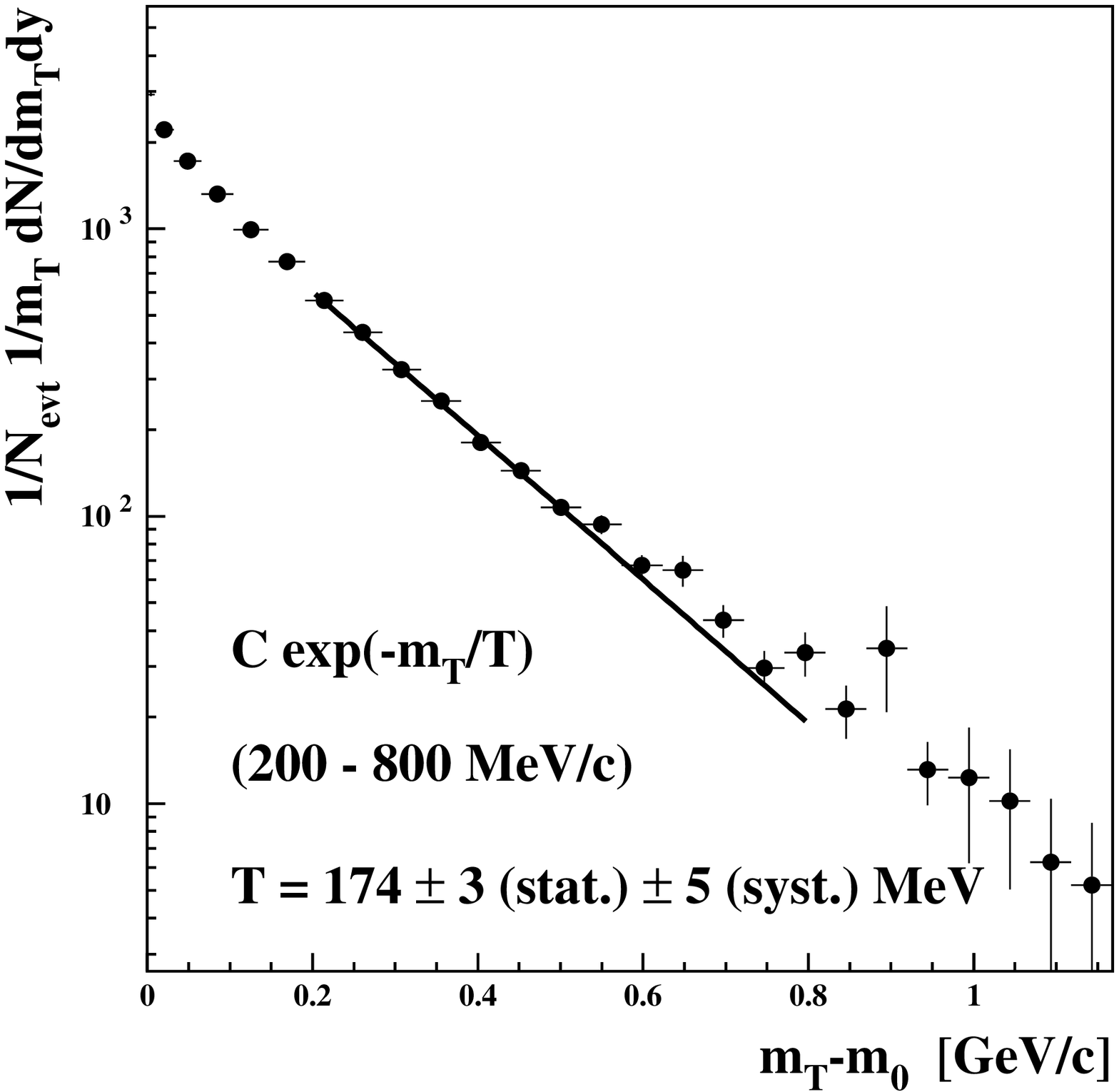}
}
\caption{$m_T$ distribution (the fit is explained in the text).}
\label{fig:figure3}
\end{figure}

The $m_T$=$\sqrt{m_\pi^2+p_T^2}$ distribution of identified $\pi^-$, averaged
over the rapidity acceptance, is shown in Fig.~\ref{fig:figure3}.
The data were corrected for geometrical acceptance and efficiency
of the chamber-camera-Time of Flight system using a full
simulation of the experimental setup.
The parameters of the simulation were optimized in an iterative 
way by comparing various distributions with the real data.
The simulated data were then treated exactly like the real data.
The measured $1/m_T \,dN/dm_T\:dy$
distribution was then fitted to the form $C$exp$(-m_T/T)$, expected for
a source in thermal equilibrium
\cite{schne}.
Such fits were applied to the data for different ranges of $m_T$,
such as the one shown in Fig.~\ref{fig:figure3}. These fits do not reproduce
the overall concave shape of the data, which
is partly due to particles originating in resonance
decays and could also be an indication of transverse flow \cite{agga}.
The shape of the $\pi^-$ $m_T$ distribution was found to be in good agreement
with that of $\pi^0$ obtained in the lead-glass calorimeter \cite{Peitz}.

\section{One-dimensional interferometry analysis}

\begin{figure}[!h]
\hspace*{4.0cm}
\resizebox{0.471\textwidth}{!}{%
  \includegraphics{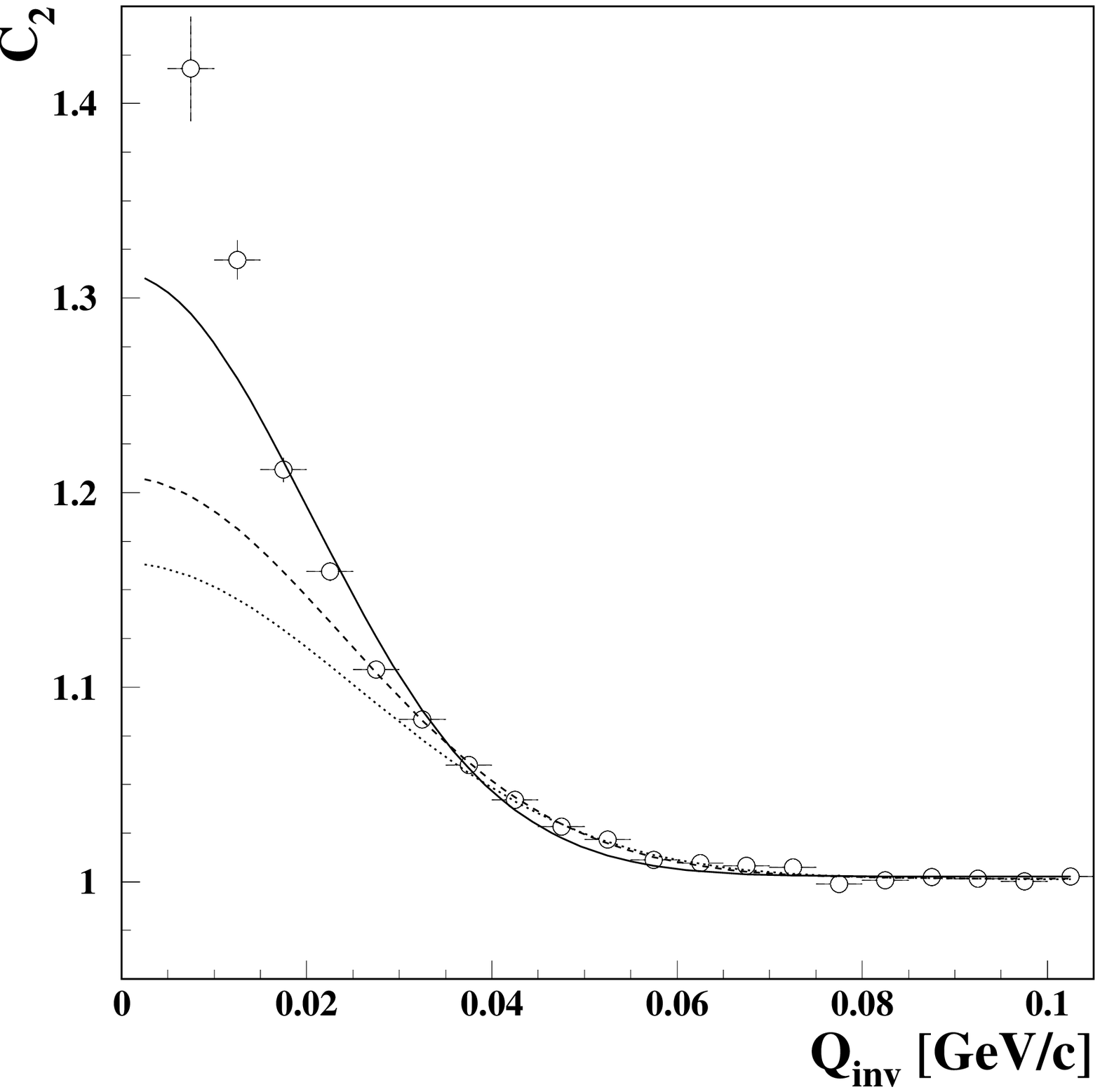}
}
\caption{$Q_{inv}$ distribution with Gaussian fits. The full line is a 
 fit made with all the points. The dashed  line is a fit for $Q_{inv}
 \geq 25$ MeV/c and the dotted line is a fit for $Q_{inv}
 \geq 40$ MeV/c. }
 \label{fig:figure4}
\hspace*{4.0cm}
\resizebox{0.471\textwidth}{!}{%
  \includegraphics{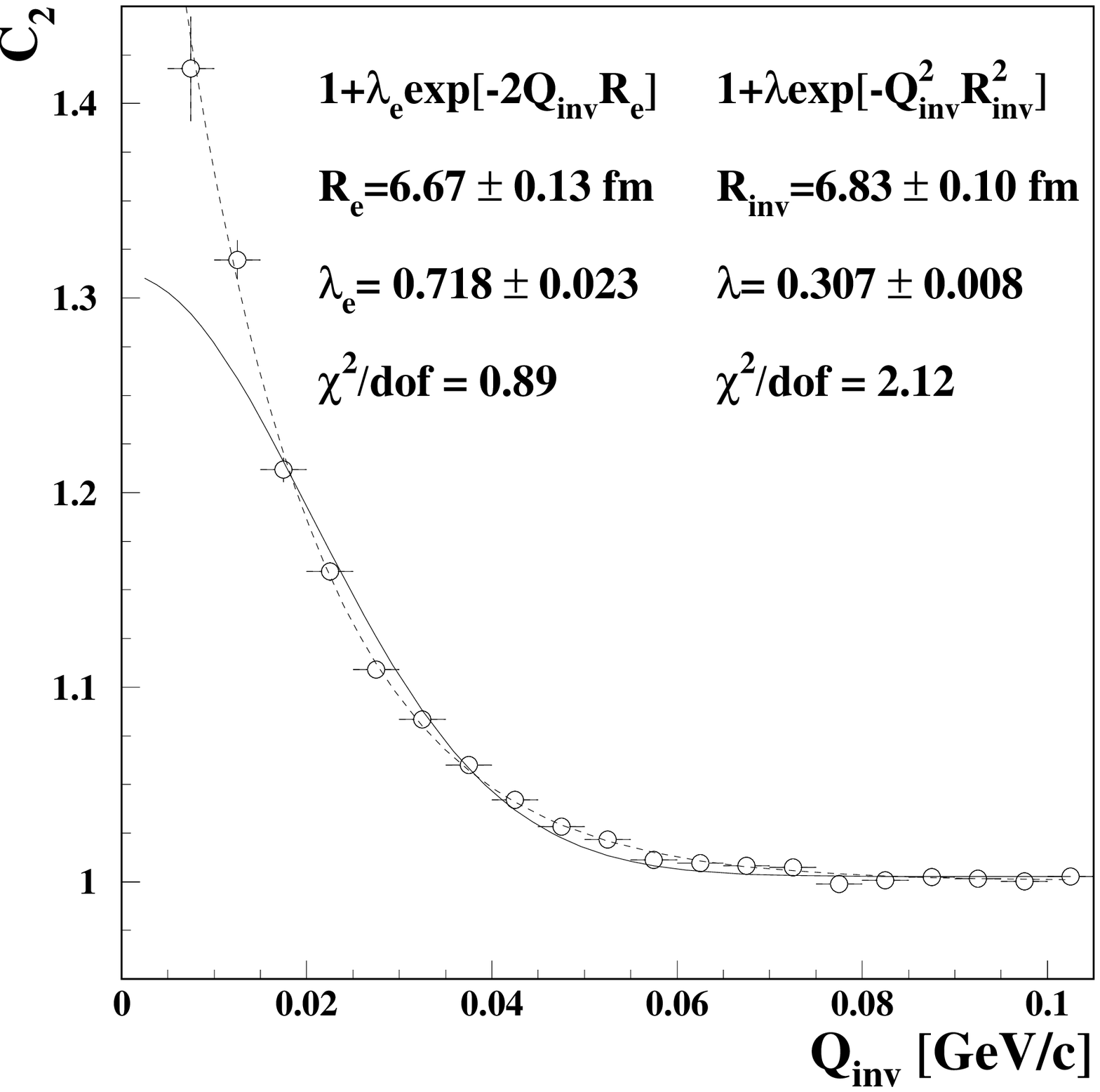}
}
\caption{$Q_{inv}$ distribution with the Gaussian fit (full line) and
 the exponential fit (dashed line). }
\label{fig:figure5}
\end{figure}

\begin{figure*}[!h]
\resizebox{1.0\textwidth}{!}{%
  \includegraphics{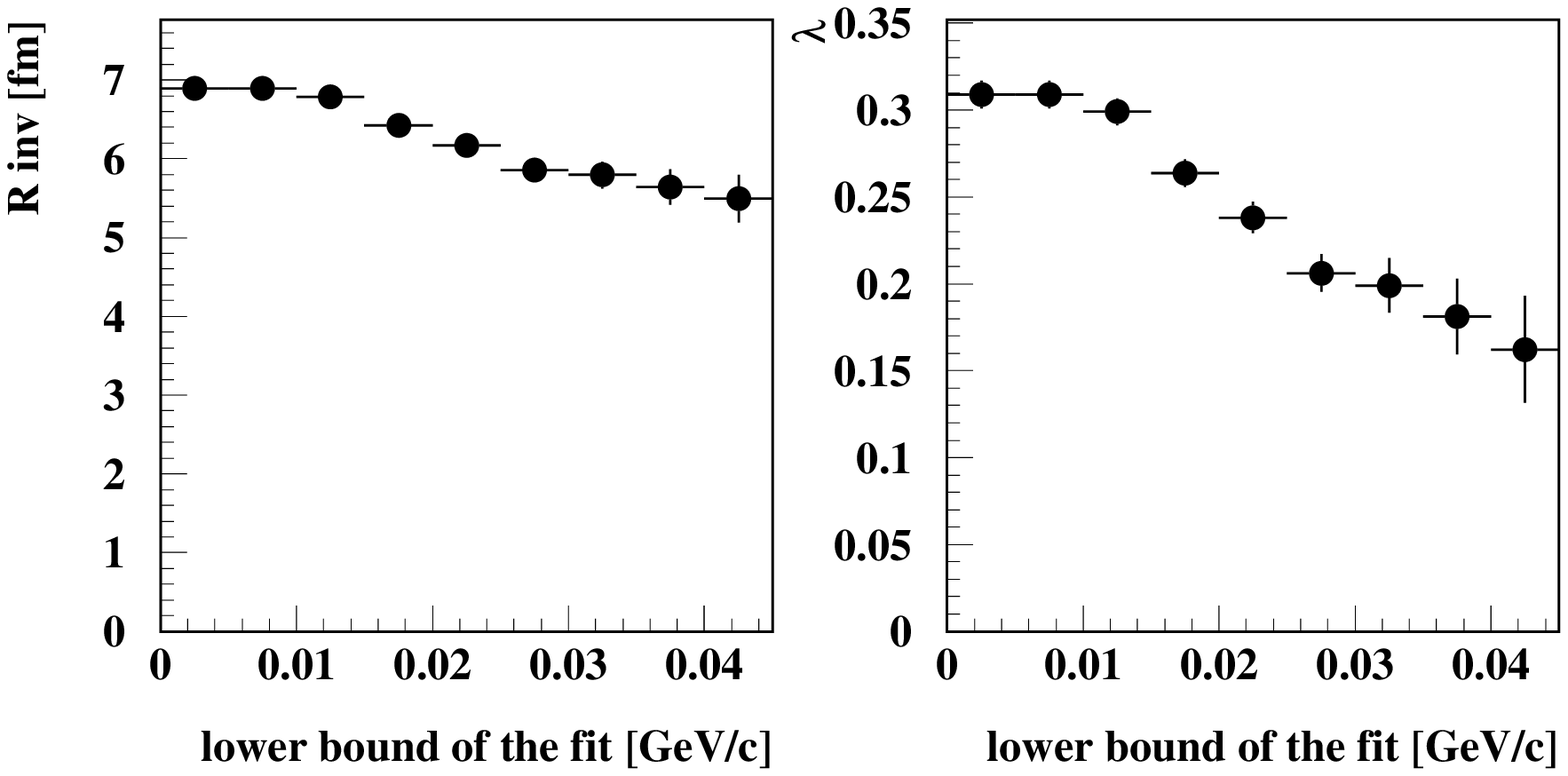}
}
\vspace*{-8.3cm}
\caption{Gaussian fit stability.}
\label{fig:figure6}
\resizebox{1.0\textwidth}{!}{%
  \includegraphics{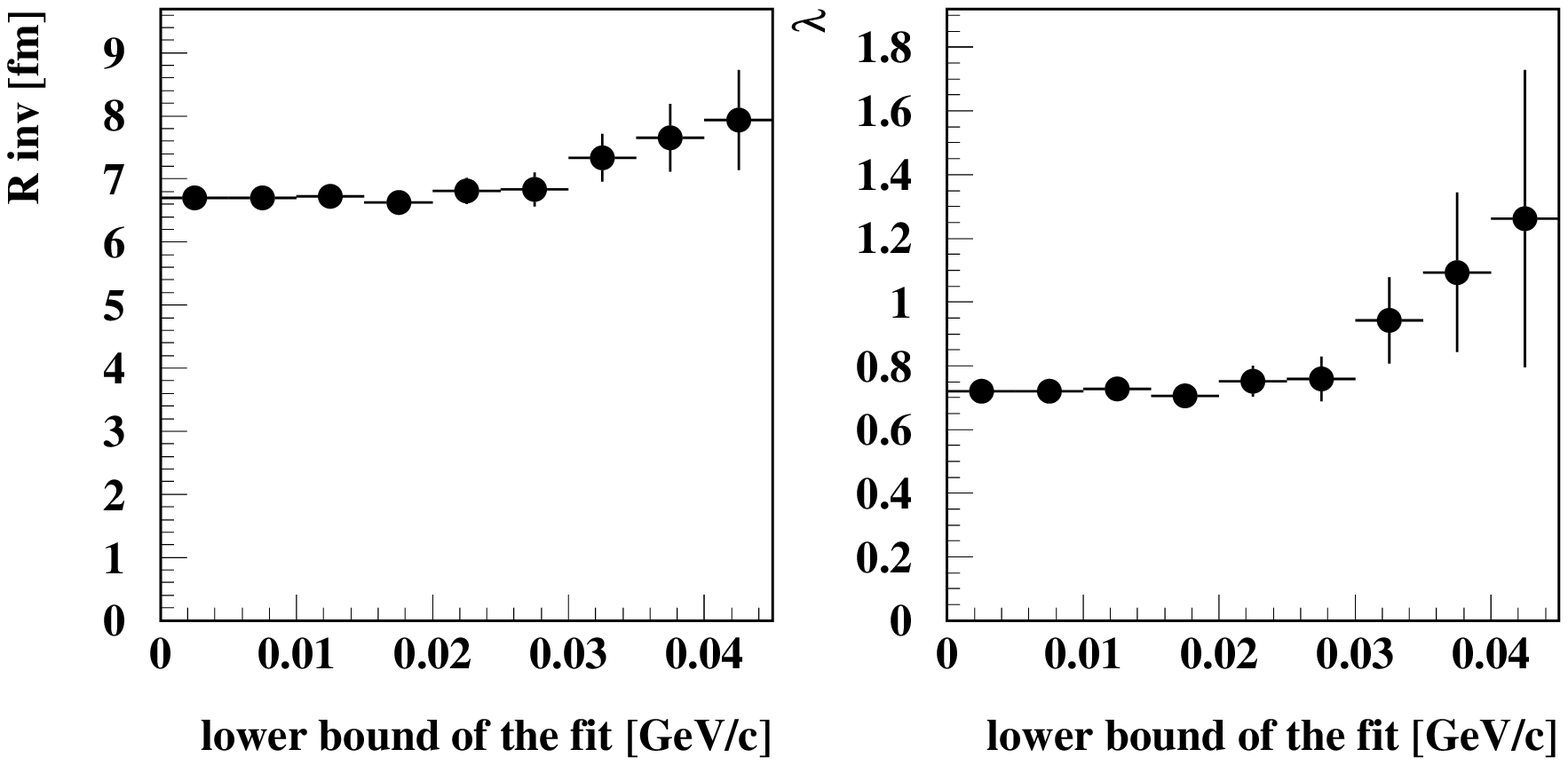}
}
\vspace*{-8.3cm}
\caption{Exponential fit stability.}
\label{fig:figure7}
\end{figure*}

For the Bose-Einstein correlation studies, the data were Coulomb corrected
in an iterative way \cite{Pratt}. The Gamow
correction was abandoned as it overcorrects the data for $Q_{inv}$
in the range of 0.1 to 0.3 GeV/c.
A fit of the form
$1+\lambda \mbox{exp} [-Q^2_{inv}R^2_
{inv}]$ was made to the $Q_{inv}$ correlation function yielding
$R_{inv}=6.83\pm0.10$ fm and $\lambda=0.307\pm0.008$.
An expanded view of the correlation distribution (Fig.~\ref{fig:figure4}) shows that the
Gaussian fit used (full line) is not perfect,
especially in the $Q_{inv}$ range of 40 to 80 MeV/c where the
tail of the experimental distribution shows an excess which 
is not well reproduced by the fit. In addition to this Gaussian fit made
over the whole range of $Q_{inv}$, Fig.~\ref{fig:figure4} shows
also different Gaussian fits using data in the ranges 25 MeV/c $\leq$
$Q_{inv}$ $\leq$ 200 MeV/c (dashed line) and 40 MeV/c $\leq$ $Q_{inv}$
$\leq$ 200 MeV/c (dotted line). These fits do not coincide.
Different radii
are then obtained for different starting points of the fit
because the shape of the distribution is not Gaussian.
This effect is independent of the severity of the track selection,
and is therefore not due to spurious tracks. This is 
summarized in Fig.~\ref{fig:figure6} where $R_{inv}$ and the corresponding $\lambda$ are
plotted as a function of the lower bound of the fit. There is a statistically
 significant drop when using a Gaussian fit. A similar behaviour is observed
when, instead of $Q_{inv}$,  $Q_{space}=\sqrt{Q^2_x+Q^2_y+Q^2_z}$ is used,
calculated in the longitudinally comoving system (LCMS) and
fitted with 1+$\lambda$exp[-$Q^2_3 R^2_3]$.
This method of fitting in varying ranges has a good sensitivity
to the shape. It has been repeated by replacing the Gaussian fit
by an exponential fit of the form
1+$\lambda_e$exp[-2$Q_{inv}R_e]$ where the factor 2 is added to
make the radius $R_e$ more comparable with $R_{inv}$.
The results (Fig.~\ref{fig:figure7}) show that the stability is better with the exponential fit.
Fig.~\ref{fig:figure5} directly compares the Gaussian
and exponential fits for $Q_{inv}$. Although the Gaussian
fit still gives an acceptable $\chi^2/$d.o.f., the exponential
fit is better everywhere. A similar conclusion is reached
when the first data point is excluded from the fit.
This result is not based on the first bins which might be more affected
by systematics due to large Coulomb correction or noise correlated with
the true track signals in the chambers. It is rather based on the high
statistics tail of the distribution which contributes in a different
way to a Gaussian or an exponential fit.
This quasi exponential behaviour
is expected by different models including
resonance decays \cite{Wied}. As a consequence small acceptance
experiments may obtain a larger radius if a Gaussian fit is
used because they are less sensitive to the tail. On
the contrary large acceptance experiments have higher statistics
at large $Q$-values, and the Gaussian fit will yield
lower values of the radius.

\section{Multi-dimensional interferometry analysis}

The multi-dimensional analysis has been done with Gaussian fits
to allow comparison with other experiments. Two different 
parameterizations have been used in the LCMS:
 
a) The standard fit in the 3-dimensional space of momentum
differences $Q_{TS}$ (perpendicular to the beam axis and to
the transverse momentum of the pair), $Q_{TO}$ (perpendicular
to the beam axis and parallel to the transverse momentum of 
the pair), and $Q_{L}$ (parallel to the beam axis) \cite{stand}.
The fitted formula
$$C_2=1+\lambda\exp[-Q_{TS}^2R_{TS}^2-Q_{TO}^2R_{TO}^2-Q_L^2R_L^2-2Q_{TO}Q_LR^2_{out-long}]$$
includes a cross term in $Q_{TO}Q_L$ as predicted \cite{Cross}.
 
b) The generalized Yano-Koonin (GYK) fit \cite{GYK} in the $Q_0$ (energy
difference of the pair), $Q_T$,$Q_L$ space according to
$$C_2=1+\lambda\exp[-Q^2_TR^2_T+(Q^2_0-Q^2_L)R^2_4-(Q{\cdot}U)^2(R^2_0
+R^2_4)]$$
where $U=\gamma(1,0,0,v_L)$, $\gamma=1/\sqrt{1-v_L^2}$ with $v_L$ in units of
c=1.
 
In the GYK approach, the radius
parameters remain invariant under longitudinal Lorentz boost, the
parameter $v_L$ connecting the ``arbitrary'' measurement frame (LCMS)
to the Yano-Koonin frame.
In addition, the extraction of the duration of emission, $R_0$,
is straightforward.
 
The consequence of the finite resolution in the measurement of the $Q$
variables
is an underestimate of the radii and $\lambda$ parameters.
Morever, as the resolution is different for each $Q$
variable, this causes a bias which varies from parameter to
parameter, leading to errors in the interpretation of the results
in a multi-dimensional analysis.
It is therefore essential to take into account the
effect of the resolution in the fitting procedure.
One way to do this is to replace the formula $C_2(\vec{Q})$
used to fit the data by
$$C_2^{rc}(\vec{Q})=\int\!\!\int\!\!\int r(\vec{Q},\vec{Q'})\: C_2(\vec{Q'})
\:d\vec{Q'}$$
which is the convolution of $C_2(\vec{Q})$ with the resolution
function $r(\vec{Q},\vec{Q'})$.
The resolution function is chosen to be Gaussian:
$$\hspace{-3.0cm}r(\vec{Q},\vec{Q'})=1/(2\pi)^{3/2}\:1/|V|^{1/2}\: \exp[-1/2\:(\vec{Q}-\vec{Q'})^T\:V^{-1}\:(\vec{Q}-\vec{Q'})]$$
The diagonal elements of the covariance matrix $V$ are equal
to the square of the resolution of the different $Q$
variables and are estimated separately as a function of
$k_T=|\vec p_{T1}+\vec p_{T2}|/2$ of  the pairs.
The non-diagonal elements are neglected.
For the one-dimensional Gaussian fit case with $\vec{Q}=Q_{inv}$,
the resolution corrected values of the fitted parameters are
$R_{inv}=7.30\pm0.12$ fm and $\lambda=0.328\pm0.009$.
\begin{table*}[!hbt]
\caption{3-dimensional analysis}
\label{tab:table1}
\begin{center}
\begin{tabular}{|lcr@{$\pm$}l|lcr@{$\pm$}l|}
\hline
\multicolumn{4}{|c|}{Standard fit in LCMS}
& \multicolumn{4}{c|}{Generalized Yano-Koonin fit}
\\
\hline
$R_{TS}$ &=& 6.41&0.13 fm& $R_T$ &=& 6.54&0.11 fm\\
$R_{TO}$ &=& 6.60&0.16 fm& $R_0$ &=& 0.01&0.69 fm\\
$R_{L}$ &=& 7.50&0.18 fm& $R_4$ &=& 7.51&0.18 fm\\
$\lambda$ &=& 0.350&0.010 & $\lambda$ &=& 0.325&0.009 \\
$R^2_{out-long}$ &=& -1.0&1.3  fm$^2$& $v_L$ &=& 0.03&0.05 \\
$\chi^2/$d.o.f.&=&\multicolumn{2}{l|}{ \enspace 1.06}
&$\chi^2/$d.o.f.&=&\multicolumn{2}{l|}{ \enspace 1.02} \\
\hline
\end{tabular}
\end{center}
\end{table*}
 
\begin{figure*}[!h]
\resizebox{1.0\textwidth}{!}{%
  \includegraphics{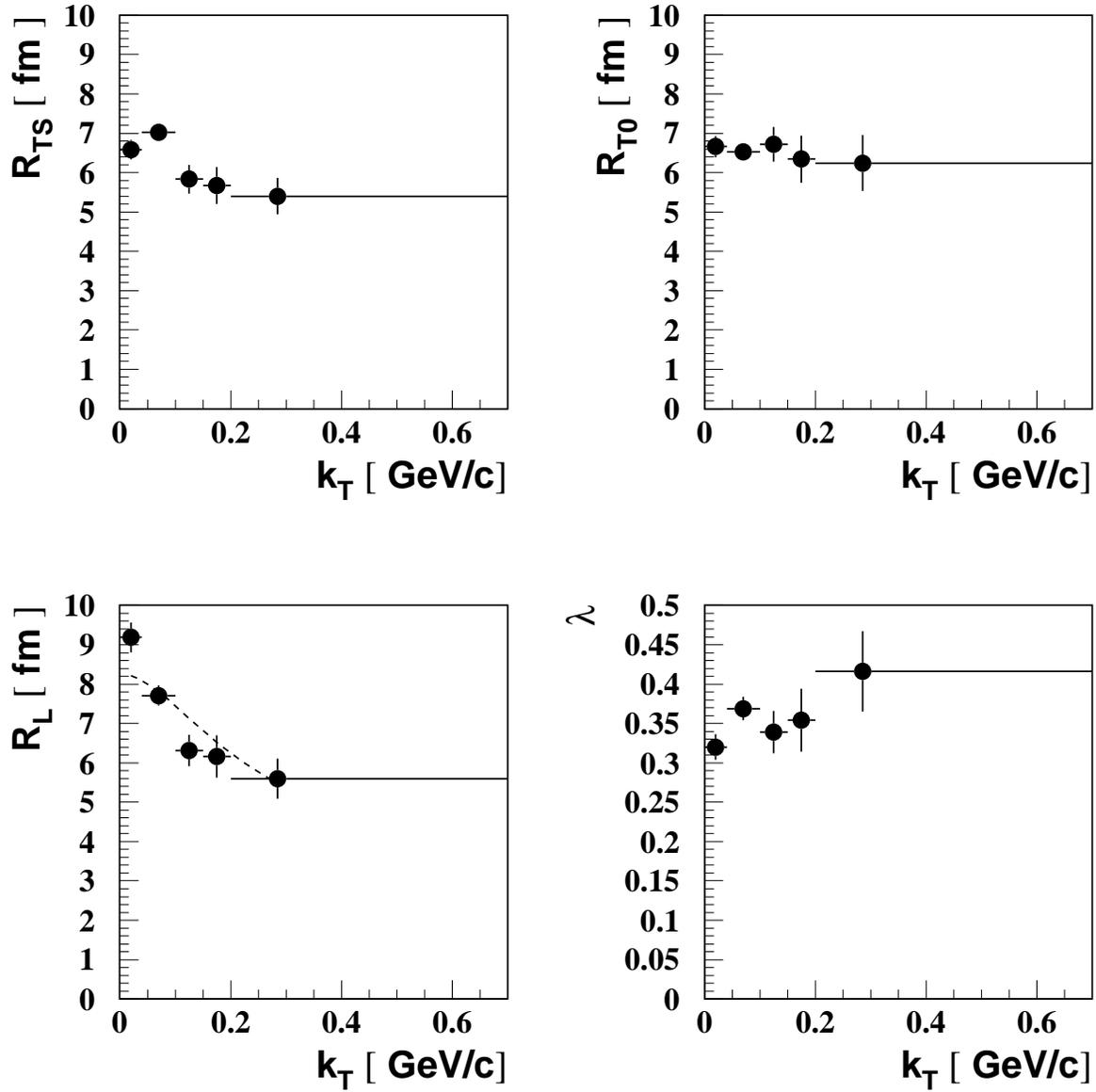}
}
\vspace*{-0.4cm}
\caption{5-parameter fit as a function of $k_T$.
The horizontal bars indicate the bin width and
the points are plotted at the average $k_T$ of the bin. The dashed
line in the bottom left figure shows the result of a fit explained
in the text.}
\label{fig:figure8}
\end{figure*}

\begin{figure*}[!h]
\vspace*{-2.0cm}
\resizebox{1.0\textwidth}{!}{%
  \includegraphics{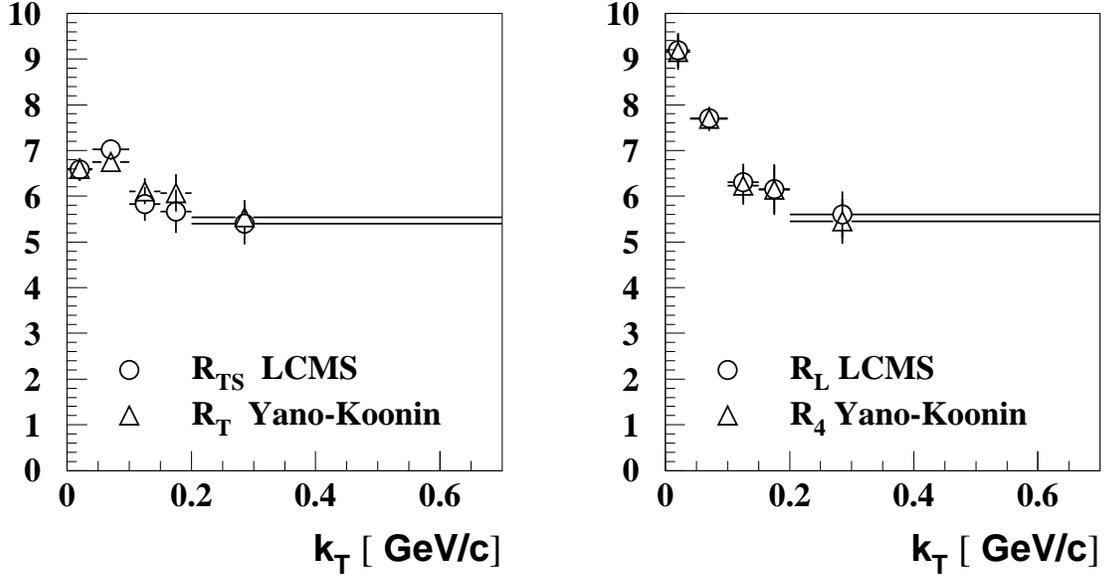}
}
\vspace*{-8.4cm}
\caption{Comparison of the $R_{TS}$ and $R_{L}$
parameters from the standard fit respectively with $R_T$ and 
$R_4$ from the GYK fit as a function of $k_T$.}
\label{fig:figure9}
\end{figure*}

\begin{figure*}[!h]
\vspace*{-2.5cm}
\resizebox{1.0\textwidth}{!}{%
  \includegraphics{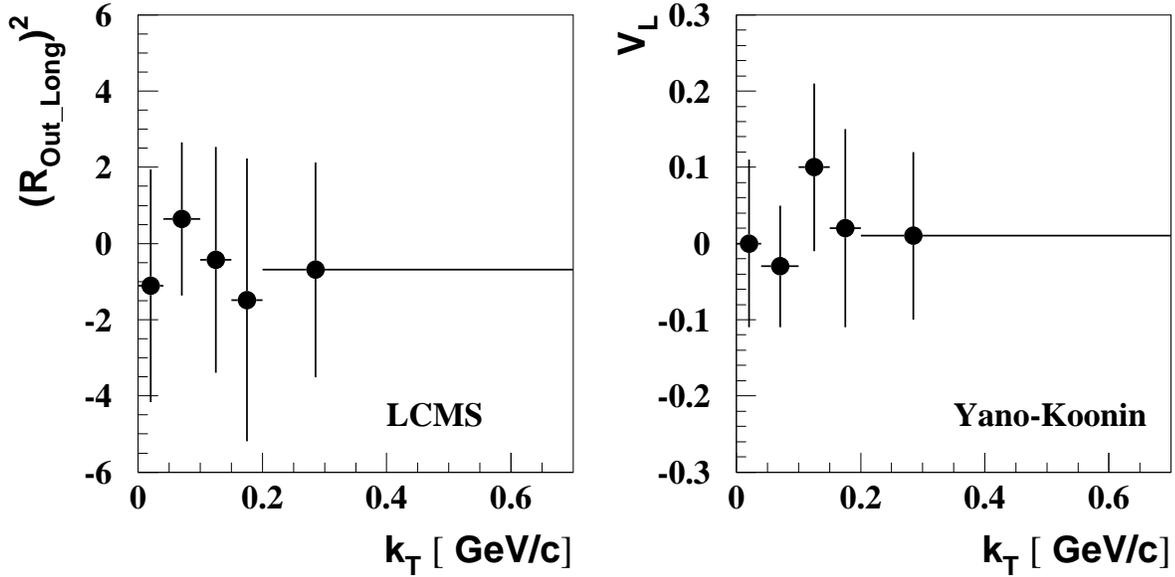}
}
\vspace*{-8.4cm}
\caption{The cross term $R^2_{out-long}$
from the standard fit and $v_L$ from the GYK fit,
both estimated in the LCMS, are compatible with 0 for
all $k_T$ bins.}
\label{fig:figure10}
\end{figure*}

The results of the multi-dimensional fits are presented in
Table~\ref{tab:table1} for
the full 1995 data sample.
A multi-dimensional analysis as a function
of $k_T$, both
with the standard 5-parameter fit and with the GYK fit is shown
in Figs.~\ref{fig:figure8}, \ref{fig:figure9}, and \ref{fig:figure10}.


 
The $R_{TS}$ and $R_{L}$
parameters from the standard fit are found to be compatible
respectively with $R_T$ and $R_4$ from the GYK fit. The cross term
$R^2_{out-long}$ from the standard fit and $v_L$ from the GYK
fit are compatible with 0. In a source undergoing a boost invariant expansion,
the local rest frame coincides with the LCMS. Both the
cross term and $v_L$
estimated in the LCMS are then expected to vanish \cite{GYK}. As this
is the case, it suggests that the source seen within the acceptance of
the experiment has a strictly boost invariant expansion.
The strong decrease of the longitudinal radius $R_L$ or $R_4$ with
$k_T$ compared to the behaviour of the transverse radii $R_T$,
$R_{TS}$, $R_{TO}$ suggests a longitudinal expansion larger than
the lateral expansion.
The longitudinal radius $R_L$ is shown with a
fit of the form 1/$\sqrt{m_T}$ with ${m_T}$=$\sqrt{m_\pi^2+k_T^2}$
inspired by the hydrodynamical expansion model.
Using $R_L=\tau_0\sqrt{T_0/m_T}$ with a freeze out
temperature $T_0$ of 120-170 MeV/c, we may extract a freeze out time
$\tau_0$ in the range of 7.5-8.9 fm/c.
Finally, the $R_0$ parameter from the GYK fit, which reflects the duration
of emission, is compatible with 0 for all $k_T$ bins, excluding a long-lived
intermediate phase.
 
Two other experiments, NA49 and NA44, have studied
charged particle interferometry in Pb+Pb collisions at
CERN energies.
The WA98 analysis is in good agreement with the NA49
results\cite{na49},
when the comparison is made for the same mean $y$ range
of 2.70, although WA98 has used identified $\pi^-$
while the NA49 analysis used unidentified negative particles.
Only the $R_0$ parameter tends to be smaller in WA98.
The direct comparison with the NA44 experiment is
not possible because NA44 and WA98 do not have the same
$y$ range.
The smaller radii measured by NA44\cite{na44}
can be explained by the larger $y$ range of its
acceptance (3.1$<y<$4.1).
 
\section{Conclusion}

The analysis of the two-particle correlation of identified $\pi^-$
from central Pb+Pb collisions at 158 AGeV gives fitted
radii of about 7 fm. This should be compared to the equivalent rms
radius of the initial Pb nucleus of 3.2 fm, which indicates a large
final state emission volume.
 
The one-dimensional correlation functions analyzed in terms of $Q_{inv}$
or $Q_{space}$ are not Gaussian. They
are better represented by exponentials. This study is based on the tail
of the distributions and not on the first bins which might be
subject to systematic effects.
One possible explanation is that this behaviour is due to
resonance effects.
Fitting Gaussians to these correlation
functions may produce different results depending on the
acceptance of the experimental setup.
 
The generalized Yano-Koonin analysis gives similar results to within the error
bars as the standard 3-dimensional analysis in the LCMS.
 
The cross term $R^2_{out-long}$ is found to be compatible with 0 in the LCMS
and the same is true of $v_L$ in the GYK fit. This suggests
that the source
undergoes a boost invariant expansion.
 
A clear dependence of the longitudinal radius parameter on $k_T$ is
observed, suggesting a larger longitudinal than transverse expansion
of the source.
In addition the short duration of emission disfavours any
long-lived intermediate phase.

\vspace{5 mm}
\hspace{-5.5 mm}
{\normalsize{\textbf {Acknowledgements}}}
\vspace{3 mm}
\\
\hspace*{4 mm}We would like to thank the CERN-SPS accelerator crew
for providing an excellent lead beam and 
the Laboratoire National Saturne for the loan of the magnet Goliath.
 
This work was supported jointly by the German BMBF and DFG, the U.S. 
DOE, the Swedish NFR, the Dutch Stichting FOM, the Swiss National Fund,
the Humboldt Foundation, the Stiftung f\"{u}r deutsch-polnische
Zusammenarbeit, the Department of Atomic Energy, the Department
of Science and Technology and the University Grants Commission of
the Government of India, the Indo-FRG Exchange Programme, the PPE
division of CERN, the INTAS under contract INTAS-97-0158, the
Polish KBN under the grant 2P03B16815, and ORISE.
ORNL is managed by Lockheed Martin Energy Research Corporation under
contract DE-AC05-96OR22464 with the U.S. Department of Energy.

\end{document}